\documentclass{article}
\usepackage{spconf}

\usepackage{amsmath, bbm}
\usepackage{amssymb}
\usepackage{amsfonts}
\usepackage{amsthm}
\usepackage{algorithm}
\usepackage{algorithmic}
\usepackage{nicefrac}
\usepackage{bm}
\usepackage{hyperref}
\usepackage{cite}
\usepackage[utf8]{inputenc} 
\usepackage[T1]{fontenc}
\usepackage{graphicx}
\usepackage[font=small]{caption}
\usepackage{tikz}
\usetikzlibrary{positioning}
\usepackage{pgfplots}
\usepackage{epstopdf}
\usepackage[labelformat=simple]{subcaption}
\usepackage{multirow}
\usepackage{lipsum}
\usepackage{mathtools}

\long\def\comment#1{}

\newtheorem{Problem}{Problem}

\input{mysymbol.sty}

\title{Distributed Link Sparsification for Scalable Scheduling Using Graph Neural Networks}

\name{Zhongyuan Zhao$^\star$, Ananthram Swami$^\dag$, and Santiago Segarra$^\star$
\thanks{Research was sponsored by the Army Research Office and was accomplished under Cooperative Agreement Number W911NF-19-2-0269. 
		The views and conclusions contained in this document are those of the authors and should not be interpreted as representing the official policies, either expressed or implied, of the Army Research Office or the U.S. Government. 
		The U.S. Government is authorized to reproduce and distribute reprints for Government purposes notwithstanding any copyright notation herein.
		\newline
		Emails: \{zhongyuan.zhao, segarra\}@rice.edu, ananthram.swami.civ@army.mil}}
\address{$^\star$Rice University, USA  \hspace{1cm} $^\dag$US Army’s DEVCOM Army Research Laboratory, USA}

\begin{document}
\ninept
\renewcommand{\baselinestretch}{0.95}
\maketitle
\begin{abstract}
Distributed scheduling algorithms for throughput or utility maximization in dense wireless multi-hop networks can have overwhelmingly high overhead, causing increased congestion, energy consumption, radio footprint, and security vulnerability.
For wireless networks with dense connectivity, we propose a distributed scheme for link sparsification with graph convolutional networks (GCNs), which can reduce the scheduling overhead while keeping most of the network capacity. 
In a nutshell, a trainable GCN module generates node embeddings as topology-aware and reusable parameters for a local decision mechanism, based on which a link can withdraw itself from the scheduling contention if it is not likely to win.
In medium-sized wireless networks, our proposed sparse scheduler beats classical threshold-based sparsification policies by retaining almost $70\%$ of the total capacity achieved by a distributed greedy max-weight scheduler with $0.4\%$ of the point-to-point message complexity and $2.6\%$ of the average number of interfering neighbors per link. 
\end{abstract}
\section{Introduction}\label{sec:intro}
\begin{figure}
    \centering
    \vspace{-0.1in}
    \hspace{0.1in}
    \begin{subfigure}[b]{0.45\linewidth}
    \includegraphics[height=1in]{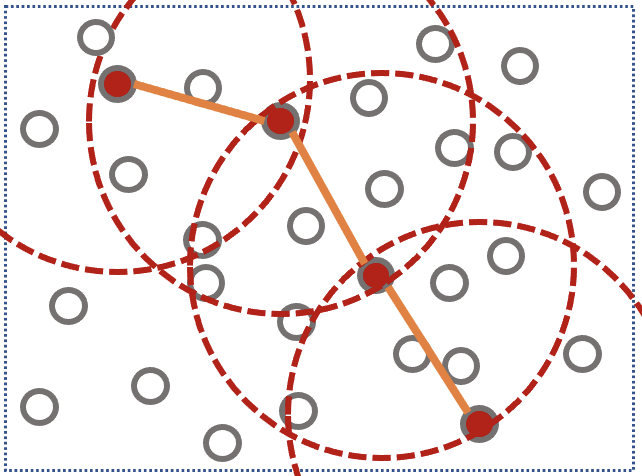}
    \caption{}
    \label{fig:motivation:dense}
    \end{subfigure}\hspace{0.1in}
    \begin{subfigure}[b]{0.45\linewidth}
    \includegraphics[height=1in]{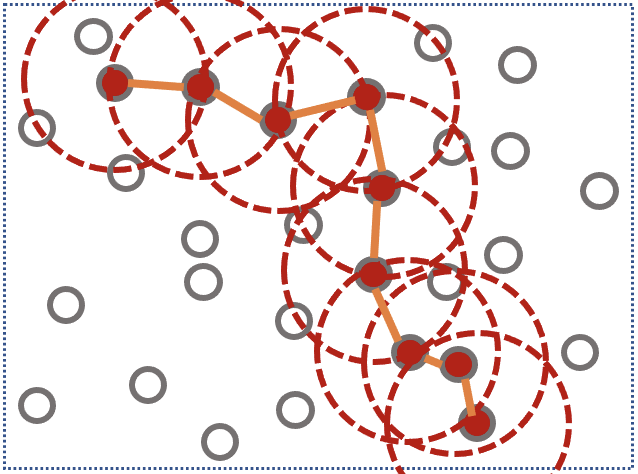}
    \caption{}
    \label{fig:motivation:sparse}
	\end{subfigure}
    \vspace{-0.15in}
    \caption{Multi-hop wireless networks with neighborhoods of different sizes. (a) Dense network with large neighborhoods. (b) Sparse network with small neighborhoods. 
    The overhead, collision, and starvation of scheduling increases with the neighborhood size, while the relays and thus overhearing increase in sparser networks.
    }
    \label{fig:motivation}
    \vspace{-0.2in}
\end{figure}

The proliferation of wireless devices and emerging machine-to-machine (M2M) traffic \cite{cisco2020}  bring new requirements to wireless networks, such as massive access, ultra-dense networks, better spectrum and energy efficiencies \cite{kott2016internet,akyildiz20206g,chen2021massive}.
 Wireless multi-hop communications is a promising solution for applications such as military communications, vehicular/UAV networks, wireless backhaul for xG, and Internet of Things (IoT) \cite{Lin06,sarkar2013ad,kott2016internet,akyildiz20206g,chen2021massive}.
\comment{As a promising solution, wireless multi-hop networks can support applications in which the infrastructure is infeasible or overloaded, such as military communications, vehicular/UAV networks, wireless backhaul for 5G and beyond,  and Internet of Things (IoT) \cite{Lin06,sarkar2013ad,kott2016internet,akyildiz20206g,chen2021massive}.}
A fundamental problem in wireless multi-hop networks is distributed resource allocation without the help of infrastructure.
This includes link scheduling, which determines which links should transmit and when should they transmit
\cite{Joo09,marques2011optimal}. 
The typical formulation of optimal scheduling in wireless multi-hop networks with orthogonal access is solving a maximum weighted independent set (MWIS) problem on a conflict graph~\cite{basagni2001finding,Joo09,cheng2009complexity,joo2012local,joo2015distributed,marques2011optimal,du2016new,paschalidis2015message,joo2010complexity,zhao2021icassp,zhao2021jstsp},
in which a vertex represents a link in the wireless network, an edge captures the interference relationship between two links, and the vertex weight is the utility of scheduling the corresponding link.
The MWIS problem is known to be NP-hard~\cite{cheng2009complexity,joo2010complexity}, and many approximate solutions have been proposed. Common approaches include distributed greedy algorithms \cite{joo2012local,joo2015distributed}, carrier sense multiple access (CSMA) \cite{ni2010qcsma,jiang2010distCSMA}, message passing algorithms \cite{paschalidis2015message,du2016new}, and their hybrids \cite{zhao2021icassp,zhao2021jstsp}.
Being designed for throughput or utility maximization~\cite{basagni2001finding}, the existing distributed schedulers introduce contention overheads or collision rates proportional to the average size of the neighborhood in the network, as illustrated in Fig.~\ref{fig:motivation}.
This can lead to prohibitively high scheduling overhead in massive access \cite{cisco2020,chen2021massive}, as well as increased congestion, instability, energy consumption, radio footprint, and security vulnerability in network operations \cite{chen2021massive,Testi2021blind,ye2004coordinated,Santi2005topology}.

Existing approaches to the aforementioned problems include topology control with antenna beams and/or transmit power \cite{Ramanathan2004ch5,Santi2005topology,ray2016hybrid}, sleep scheduling \cite{ray2016hybrid,ye2004coordinated,guha2011greenwave,long2020collaborative}, and cross-layer optimization \cite{lin2010lowcomplexity,xiang2014energy,wu2020energy}.
Topology control seeks to maintain a connected topology with minimal energy consumption, which however would increase overhearing and latency (e.g., larger hop distance as shown in Fig.~\ref{fig:motivation:sparse}), as well as the risks of congestion and failure at some critical nodes. 
Sleep scheduling can reduce network density by putting nodes into periodic sleep mode in standalone or coordinated manner, and generally work well for low-duty-cycle devices with very low traffic demand.
Cross-layer schemes jointly optimize power control, link scheduling, and routing for energy efficiency, but they generally have the same or higher control overhead than the above-mentioned MWIS schedulers, may need centralized computing \cite{xiang2014energy,wu2020energy}, and only apply to either time-slotted or random access networks.

In this work, we consider scalable scheduling in general wireless multi-hop networks with heterogeneous devices (e.g., with different importance, air-interfaces, and tasks), orthogonal access, and high connection density, where existing approaches \cite{Ramanathan2004ch5,Santi2005topology,ray2016hybrid,ye2004coordinated,guha2011greenwave,long2020collaborative,lin2010lowcomplexity,xiang2014energy,wu2020energy} are less effective or not scalable.
Since network capacity is mainly constrained by interference (connection density) \cite{li2001capacity} and given the growing bandwidth of air-interfaces \cite{cisco2020}, we can trade off latency for connectivity by scheduling fewer transmissions each with larger payloads for each link. 
For example, a desired subset of links could be excluded from contention by a global cut-off threshold derived from the empirical cumulative distribution function (eCDF) of per-link utility in the network.
With a utility function based on queue length \cite{joo2012local,joo2015distributed} or sojourn time \cite{hai2018delay}, each link will eventually join the contention as new packets arrive or time passes.
Importantly, fewer concurrent transmission attempts will reduce scheduling overhead, collision, and idle listening.
Furthermore, this approach could also be incorporated into existing solutions \cite{Ramanathan2004ch5,Santi2005topology,ray2016hybrid,ye2004coordinated,guha2011greenwave,long2020collaborative,lin2010lowcomplexity,xiang2014energy,wu2020energy}.

In this paper, we go beyond the statistical approach discussed above, and propose a link sparsification scheme based on graph convolutional networks (GCNs), which assesses a link using both the utility value and topological information, and withholds it from a scheduling contention that it could not win. 
In a nutshell, a trainable GCN generates two topology-aware multipliers to respectively scale the per-link utility and the global cut-off threshold for each link, which drive the decision of whether or not a link should contest for scheduling.
Inspired by~\cite{zhao2021icassp,zhao2021jstsp}, the overall architecture of our sparse scheduler is composed of a GCN followed by a non-differentiable distributed contention process, which can be a distributed greedy scheduler \cite{joo2012local,joo2015distributed} for time-slotted networks, or   weighted CSMA \cite{ni2010qcsma,jiang2010distCSMA} for random access networks. 
We propose a novel two-stage approach to train our GCN for link sparsification:
1) The GCN is trained to mimic the performance of a baseline (non-sparse) greedy scheduler while reducing the number of links considered in the contention, and
2) The GCN is trained to outperform the baseline of a sparse greedy scheduler with different sparsity thresholds. 
Although our method relies on centralized training, it can be deployed in a fully distributed manner thanks to the distributed nature of the GCN and the subsequent scheduler.

\vspace{1mm}
\noindent
{\bf Contribution.} The contributions of this paper are twofold:
{1) We propose the first GCN-based distributed link sparsification scheme for wireless scheduling that exploits the topology of the interference graph, 
and 
2) Through numerical experiments, we demonstrate the superior performance of the proposed method as well as its generalizability over different topologies.}

\vspace{-2mm}
\section{System Model and Problem Statement}
\label{sec:problem}

Consider a wireless multi-hop network, where an (undirected) link $(i,j)$ implies that user $i$ and user $j$ can communicate with each other. 
A flow describes the stream of packets from a source user to a destination user, and may pass through multiple links determined by a routing scheme. 
In each link, there is a queuing system $q$ for packets of all the flows as well as exogenous arrivals.

To describe the scheduling algorithm, we define \emph{conflict graph}, $\mathcal{G}(\ccalV,\ccalE)$, as follows: a vertex $v\in\ccalV$ represents a link in the wireless network, and the presence of an undirected edge $e=(v_a,v_b)\in\ccalE$ captures the interference relationship between links $v_a, v_b \in\ccalV$, which is considered to follow a physical distance model~\cite{cheng2009complexity}. 
For example, two links interfere with each other if their incident users are within a certain distance such that their simultaneous transmission will cause the outage probability to exceed a prescribed level, or they share the same user with only one radio interface.
For the rest of this paper, we focus on the conflict graph $\mathcal{G}$, which we assume to be known; see, e.g.,~\cite{yang2016learning} for its estimation. 
In principle, the interference zone of each link (hence $\mathcal{G}$) depends on the transmit power and antenna directivity of the corresponding users.
To avoid this dependency, we consider a simplified scenario in which all the users transmit at power levels that are time-invariant. 
From the definition of $\mathcal{G}$, a legal schedule must be a set of wireless links that can communicate simultaneously in time and frequency under orthogonal access, 
which forms an independent (vertex) set in $\mathcal{G}$ defined as a set of nodes with no edges connecting each other.

We describe the \emph{network state} at time $t$ by the tuple $(\mathcal{G}(t), \mathbf{u}(t))$ consisting of the conflict graph $\mathcal{G}(t)$ (potentially changing over time) and a utility vector $\mathbf{u}(t)$ collecting $u(v,t)\in\reals_{+}$ \forall $v\in\ccalV$.
The utility $u(v,t)$ can capture, e.g., the hierarchical class, backlogs, sojourn time, and link rate of the wireless link $v$.
We denote by $c(\cdot)$ the scheduling contention process (e.g., a distributed greedy scheduler) that maps every network state into an approximate MWIS of the graph.
For notational simplicity, we henceforth omit $t$ for operations in the same time slot, and denote the total utility of a vertex set $\boldsymbol{v}$ by $u(\boldsymbol{v}) = \sum_{v\in\boldsymbol{v}}u(v)$.
Since our goal is to reduce the signaling overhead in the network, we want each link to decide if it will contest for scheduling based on its own information, without talking with its neighbors.
Hence, we want to find vertex-specific functions $h_v$ for all $v \in \mathcal{V}$ such that their application to the local utilities $u(v)$ determine whether a vertex should be considered for contention or not.
Formally, we define our problem as follows.

\vspace{-1mm}
\begin{Problem}\label{P:main} 
    Given a distribution $\mathcal{N}$ over network states $(\mathcal{G}, \mathbf{u})$, we want to obtain the optimal link sparsification functions $\{h_v^*\}$ for all $v \in \mathcal{V}$ as
    \vspace{-2mm}
    \begin{subequations}\label{eq:sp}
	\begin{align}
	 \{h_v^*\} &= \argmax_{\{h_v\}} \,\,\,  \mathbb{E}_{\mathcal{N}} \Big( u(\hat{\boldsymbol{v}}^{s}) - \alpha |\ccalE^s | \Big) \label{eq:sp:obj}\\[2pt]
	\text{s.t. } \,\,\, \ccalG^{s} & = \ccalG \setminus \{v| v\in\ccalV, h_v(u(v))\leq 0\}, \;\label{eq:sp:res}\\
	\bbw^s &= [h_v(u(v))], \; \forall\; v\in \ccalV^s,\; \label{eq:sp:ws}\\
	\hat{\boldsymbol{v}}^{s} &= c(\mathcal{G}^{s}, \mathbf{w}^s). \label{eq:sp:mwis}
	\end{align} 
\end{subequations}
\end{Problem}

\vspace{0.5mm}
\noindent
To better understand Problem~\ref{P:main}, first notice that constraint~\eqref{eq:sp:res} defines the sparsified conflict graph $\ccalG^s(\ccalV^s,\ccalE^s)$ by removing from the original conflict graph $\ccalG$ those nodes with non-positive value of $h_v(u(v))$.
Thus, it is immediate that the form of functions $h_v$ has a direct influence on the level of sparsity of $\ccalG^s$.
Constraint~\eqref{eq:sp:ws} defines a modified utility vector $\bbw^s$ for the nodes of the sparsified graph $\ccalG^s(\ccalV^s,\ccalE^s)$.
Furthermore, constraint~\eqref{eq:sp:mwis} determines the scheduled vertices $\hat{\boldsymbol{v}}^{s}$ by applying a predefined scheduler $c(\cdot)$ to the sparsified graph.
In~\eqref{eq:sp:obj} we have two competing objectives: we want to maximize the utility of the scheduled vertices while minimizing the number of edges in our sparsified graph (since these determine the message complexity of the contention process).
Consequently, our objective function linearly combines both terms with a relative weighting parameter $\alpha$.
Finally, it should be noted that we do not want to find sparsifying functions for a specific network state but rather find functions that generalize well across a whole distribution of network states. 
Hence, the inclusion of an expected value in~\eqref{eq:sp:obj}.

Finding an exact solution to Problem~\ref{P:main} is extremely challenging for several reasons including:
i) The non-differentiable scheduler $c(\cdot)$ prevents direct application of gradient-based approaches,
ii) The optimization is over the space of functions, which is infinite-dimensional, and 
iii) The sparsification functions should be valid for a distribution of network states, possibly associated with conflict graphs of different sizes and topologies.
In the next section we present our solution to Problem~\ref{P:main}, which addresses these challenges.

\begin{figure*}
	\vspace{-0.1in}
	\centering
	\hspace{-0.25in}
	\begin{subfigure}[b]{0.26\linewidth}
		\includegraphics[width=\linewidth]{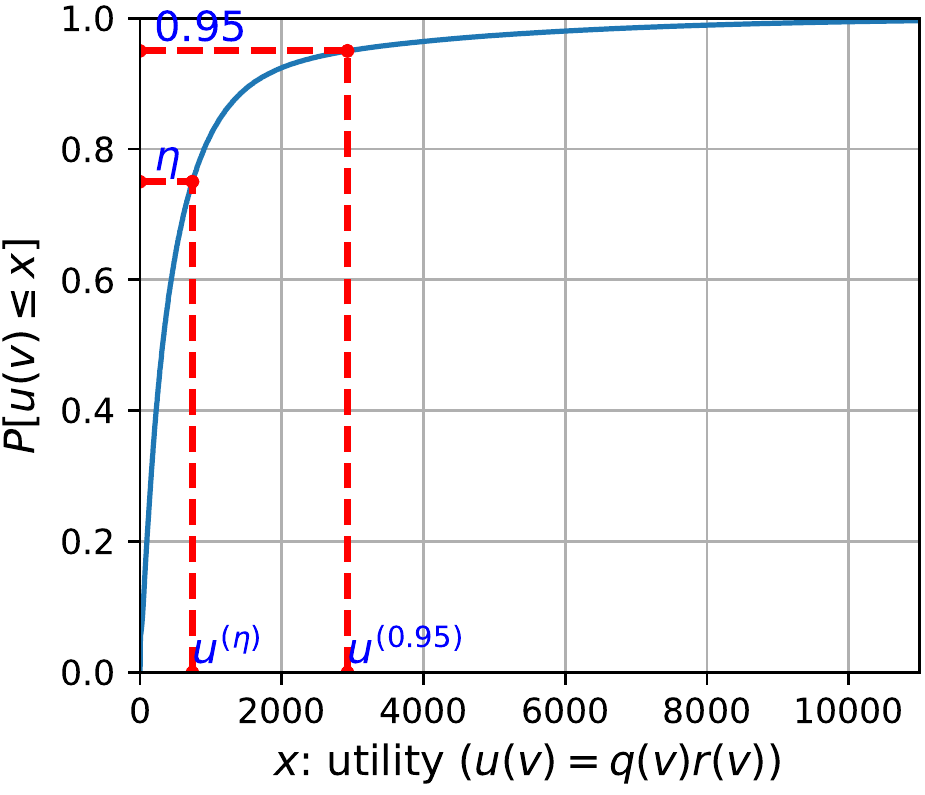}\vspace{-0.05in}
		\caption{}
		\label{fig:ecdf}
	 \end{subfigure}%
	~     
	\begin{subfigure}[b]{0.45\linewidth}
		\includegraphics[width=\linewidth]{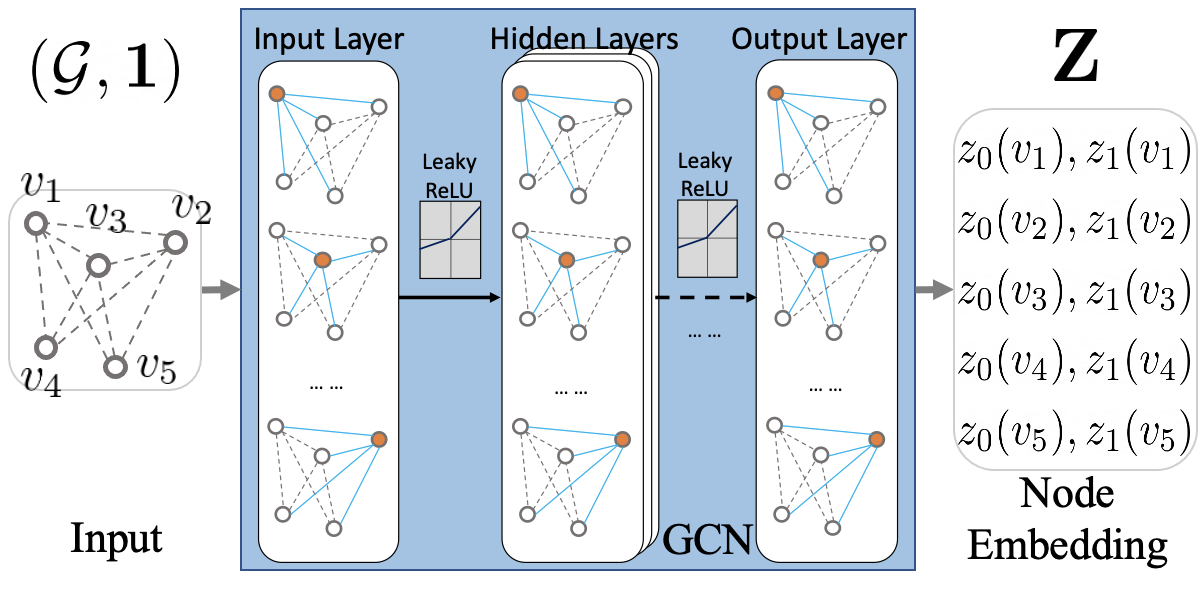}\vspace{-0.05in}
		\caption{}
		\label{fig:scheme1}
	\end{subfigure}  
	~     \hspace{0.05in}
	\begin{subfigure}[b]{0.24\linewidth}
		\includegraphics[width=\linewidth]{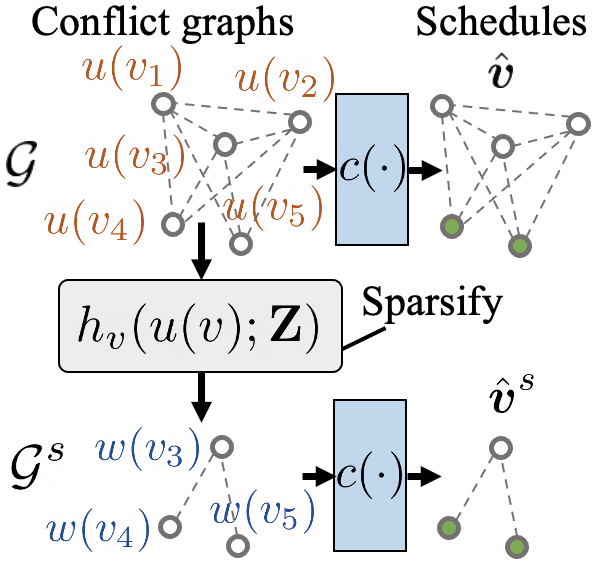}\vspace{-0.05in}
		\caption{}
		\label{fig:scheme2}
	\end{subfigure}  
	\vspace{-0.15in}
	\caption{ Architecture of GCN-based distributed sparse scheduler: (a) Select $u^{(\eta)}$ based on the eCDF of utility, (b) parameters $\bbZ$ generated by a featureless GCN based on conflict graph $\ccalG$, and (c) sparse graph $\ccalG^s$ and topology-aware utility $\bbw^{s}$ created by the parameterized functions $ \{h_v(u(v); \bbZ)\} $ based on $(\ccalG,\bbu)$, then the distributed scheduler generates schedules $\hat{\boldsymbol{v}}=c(\ccalG,\bbu)$ and $\hat{\boldsymbol{v}}^s=c(\ccalG^s,\bbw^s)$.} \label{fig:system}
	\vspace{-0.2in}
\end{figure*}

\section{Link Sparsification with GCNs}
\label{sec:solution}

In order to (approximately) solve Problem~\ref{P:main}, we restrain ourselves to a parametric family of functions $\{h_v\}$ given by
\begin{equation}\label{E:param}
h_v(u(v); \bbZ) = z_{0}(v) u(v) \, H\left(z_{0}(v)u(v)- z_{1}(v)u^{(\eta)}\right),
\end{equation}
where {$z_0(v)$ and $z_1(v)$ are parameters employed by an individual link to decide on whether or not to join the scheduling contention,} $\bbZ=\left[\bbz_0,\bbz_1\right] \in\reals^{|\ccalV|\times2}$ is a matrix collecting the link parameters $z_0(v)$ and $z_1(v)$ for all $v \in \ccalV$, 
$H(\cdot)$ is a Heaviside (step) function, and $u^{(\eta)}$ is the $\eta$-quantile utility under  network state distribution $\mathcal{N}$, as illustrated in Fig.~\ref{fig:ecdf}.
To better understand~\eqref{E:param}, notice that when $\bbZ=[\boldsymbol{1}, \boldsymbol{1}]$ the sparsification policy induced by $h_v$ boils down to a baseline thresholding method.
More precisely, the Heaviside function will compare the vertex utility $u(v)$ with the prescribed threshold $u^{(\eta)}$.
If $u(v)\leq u^{(\eta)}$, then $h_v(u(v)) = 0$ and the node is excluded from the sparse graph [cf.~\eqref{eq:sp:res}].
On the other hand, if $u(v)$ is larger than $u^{(\eta)}$, $v$ is included in the sparse graph with the same utility $h_v(u(v)) = u(v)$.
In this context, our proposed parameterization in~\eqref{E:param} is a natural extension of this classical statistical baseline.


In order to define functions $h_v$ that generalize well across different graphs $\mathcal{G}$, we define the parameters $\bbZ$ to be a function of the underlying topology.
More precisely, we propose to have $\bbZ = \Psi_{\ccalG}(\boldsymbol{1};\mathbf{\bbomega})$, 
where  $\Psi_{\ccalG}$ is an $L$-layered GCN defined on the conflict graph $\ccalG$, and $\bbomega$ is the collection of trainable parameters of the GCN.


We define the output of an intermediate $l$th layer of the GCN as $\bbX^l \in\reals^{|\ccalV|\times g_{l}}$, and $\bbX^0 = \boldsymbol{1}^{|\ccalV|\times 1}$, $\bbZ = \bbX^L $, then 
the $l$th layer of the GCN is expressed as:
\begin{equation}\label{E:gcn}
	\mathbf{X}^{l} = \sigma\left(\mathbf{X}^{l-1}{\bbTheta}_{0}^{l}+\bbcalL \mathbf{X}^{l-1}{\bbTheta}_{1}^{l}\right)\;, l\in\{1,\dots,L\}.
\end{equation}
In~\eqref{E:gcn},  
$\bbcalL$ is the normalized Laplacian of $\ccalG$, ${\bbTheta}_{0}^{l}, {\bbTheta}_{1}^{l} \in \mathbb{R}^{g_{l-1} \times g_{l}}$ are trainable parameters, and $\sigma(\cdot)$ is the activation function. 
The activation functions of the input and hidden layers are selected as leaky ReLUs whereas a linear activation is used for the output layer. 
The input and output dimensions are set as $g_{0}=1, g_{L}=2$.

The downstream architecture of the entire distributed sparse scheduling is illustrated in Figs.~\ref{fig:system}.
First, the prior knowledge of eCDF of the utility values from empirical data of network operations is collected, as shown in Fig.~\ref{fig:ecdf}, based on which a global cut-off threshold is selected as $u^{(\eta)}$.
Next, at the network level, the trained GCN observes the topology of the network and generates node embeddings as the local parameters $\bbZ$.
Then, the sparse graph $\ccalG^s$ and topology-aware utility vector $\bbw^s$ are obtained by the parameterized functions $ \{h_v(u(v); \bbZ)\} $  according to \eqref{eq:sp:res} and \eqref{eq:sp:ws}, respectively. 
Finally, the sparse schedule is obtained by the distributed scheduler as $\hat{\boldsymbol{v}}^s=c(\ccalG^s,\bbw^s)$ with lower overhead.

Since $\bbcalL$ in \eqref{E:gcn} is a local operator on $\ccalG$, $z_0(v)$ and $z_1(v)$ can be computed in a distributed manner through neighborhood aggregation at $v$ with $L$ rounds of local exchanges between $v$ and its neighbors. 
The local communication complexity (defined as the rounds of local exchanges between a node and its neighbors) of GCN is $\ccalO(L)$.
Hence, the local computational and communication costs can be controlled by modifying the number of layers $L$ in the GCN.
Notice that $\bbZ$ can be reused over time slots until the conflict graph $\ccalG$ changes.
Importantly, the constant local communication complexity is a key aspect to promote scalability, whereas the reusability of $\bbZ$ is critical for overhead reduction.

To accurately control the sparsification ratio, we scale $\bbz_1$ by its expectation as $z_1(v)\leftarrow z_1(v)/\mathbb{E}_{\ccalN}(\overline{z_1})$ after GCN, where $\mathbb{E}_{\ccalN}(\overline{z_1})$ is the expected average of $z_1(v)$ over all $v\in\ccalG$ under network state distribution $\ccalN$.
Since $\mathbb{E}_{\ccalN}(\overline{z_1})$ can be computed offline, this operation does not introduce any additional communication complexity. 

\vspace{-0.1in}
\subsection{Two-Stage Training}

The parameters $\bbomega$ in the GCN are trained on a set of random network states $(\ccalG(i),\bbu(i))$ drawn from the distribution $\ccalN$. 
To address the conflicting objectives in \eqref{eq:sp:obj}, we break the training into two stages of relatively low-complexity training schemes.
At stage 1, we employ $u^{(0.95)}$ in \eqref{E:param} and the vanilla (dense) scheduler $c(\ccalG(i), \bbu(i))$ as the baseline. 
At stage 2, we employ a random cut-off quantile $0<\eta<1$ in \eqref{E:param}, and the baseline is the statistical method, $h_v(u(v); [\boldsymbol{1}, \boldsymbol{1}])$. 
In each stage, we run the GCN-based sparse scheduler on the training dataset and collect experience tuples $\left(\ccalG(i), \bbu(i), \hat{\boldsymbol{v}}^s(i), {\boldsymbol{v}}^r(i), \bbrho_0(i), \bbrho_1(i)\right)$, for $i\in\{0,\dots,N\}$.
${\boldsymbol{v}}^r=\{v| v\in\ccalV, h_v(u(v))\leq 0\}$ is the set of removed vertices.
$\bbrho_0$ and $\bbrho_1$ are the target vectors for $\bbz_0$ and $\bbz_1$, respectively, which capture the reward signals to maximize the objective in \eqref{eq:sp:obj}.
Vector $\bbrho_0$ is defined with respect to (w.r.t.) the baseline \cite{zhao2021icassp,zhao2021jstsp}:
\begin{equation}\label{E:reward}
	\bbrho_0 = \varepsilon {\bbv}^s + \bbz_0 \odot (\boldsymbol{1}-\bbv^s),\; \varepsilon = u(\hat{\boldsymbol{v}}^s) / u(\hat{\boldsymbol{v}}^b)\;,
\end{equation}
where $\varepsilon$ is the approximation ratio (AR) for total utility, $\hat{\boldsymbol{v}}^b $ is the baseline schedule, and ${\bbv}^s$ is the indicator vector of schedule $\hat{\boldsymbol{v}}^s$.
\eqref{E:reward} encourages (discourages) schedules that are better (worse) than the baseline. 
Vector $\bbrho_1$ for stages 1 and 2 is respectively defined as:
\begin{subequations}\label{E:rho}
\begin{align}
\bbrho_1 &= \bbrho_2, \; \bbrho_2 = 
    (b/u^{(\eta)})\bbz_0 \odot\bbu \odot\bbv^r   
    + \bbz_1 \odot (\boldsymbol{1} - \bbv^r),\; 
     \label{E:rho1:p1}\\
    \bbrho_1 &= \bbrho_3/\overline{\rho_3},\;
    \bbrho_3 = \bbrho_2 - 0.2\bbz_1\odot\bbv^s,\;
    \label{E:rho1:p2}
\end{align}
\end{subequations}
where $ b= 0.9 \mathbbm{1}(\varepsilon < \delta) + 1.1 \mathbbm{1}(\varepsilon \geq \delta) $, $0<\delta<1$ is the target AR for $\varepsilon$ (e.g., $\delta=0.97$), $\bbv^r$ is the indicator vector of set ${\boldsymbol{v}}^r$, and $\overline{\rho_3}=(\boldsymbol{1}^\top \bbrho_3)/|\bbrho_3|$ is the average of the elements of vector $\bbrho_3$. 
Intuitively, \eqref{E:rho1:p1} encourages removing more vertices as long as the target AR for utility w.r.t. the dense scheduler is met in stage 1.
In stage 2, \eqref{E:rho1:p2} encourages the GCN to outperform the statistical baseline on total utility, with an average threshold equal to $u^{(\eta)}$.
This drives the GCN to consistently promote peripheral vertices under any $\eta$, thus reducing the edges in sparse graphs, i.e., smaller $|\ccalE^s|$ in \eqref{eq:sp:obj}. 
Accordingly, a root-mean-square loss is adopted to train our GCN
\begin{equation}\label{E:gcn_loss}
\ell(\bbomega; \ccalG(i), \bbu(i)) = |\ccalV|^{-\frac{1}{2}} \lVert \bbZ(i)-\left[\bbrho_0(i),\bbrho_1(i)\right] \rVert_{2}.
\end{equation}
With the loss in \eqref{E:gcn_loss} and the collected experience tuples, we update the parameters $\bbomega$ of the GCN through batch training, employing the Adam optimizer and exponentially decaying learning rates.


\section{Numerical experiments}
\label{sec:results}

\begin{figure}[t]
\centering
\vspace{-0.1in}
\begin{subfigure}[b]{0.48\linewidth}
    \includegraphics[width=\linewidth]{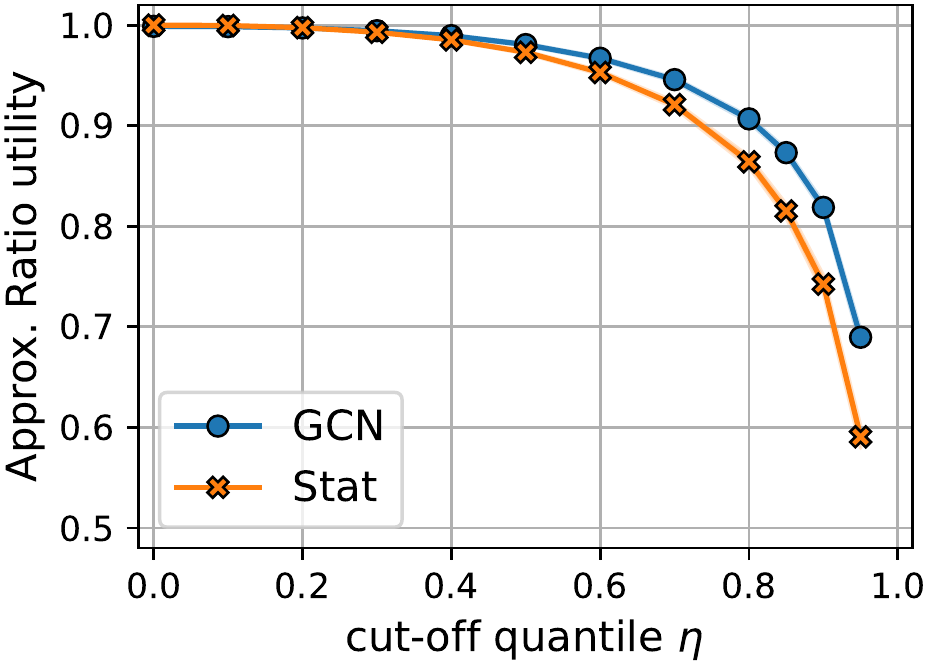}\vspace{-0.1in}
    \caption{}
    \label{fig:results:util}
\end{subfigure}
\begin{subfigure}[b]{0.48\linewidth}
    \includegraphics[width=\linewidth]{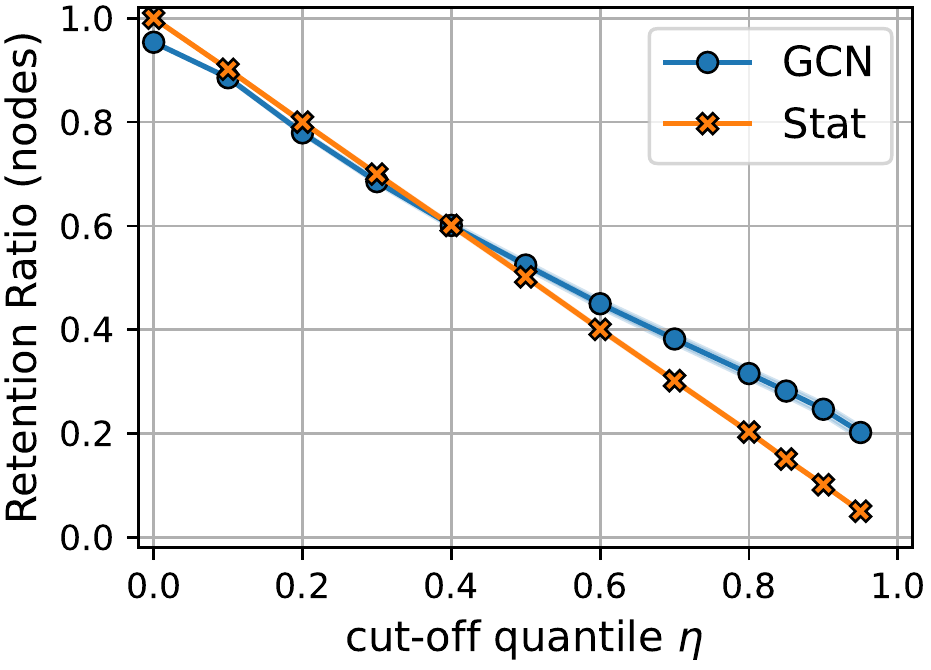}\vspace{-0.1in}
    \caption{}
    \label{fig:results:size}
\end{subfigure}\\
\begin{subfigure}[b]{0.48\linewidth}
    \includegraphics[width=\linewidth]{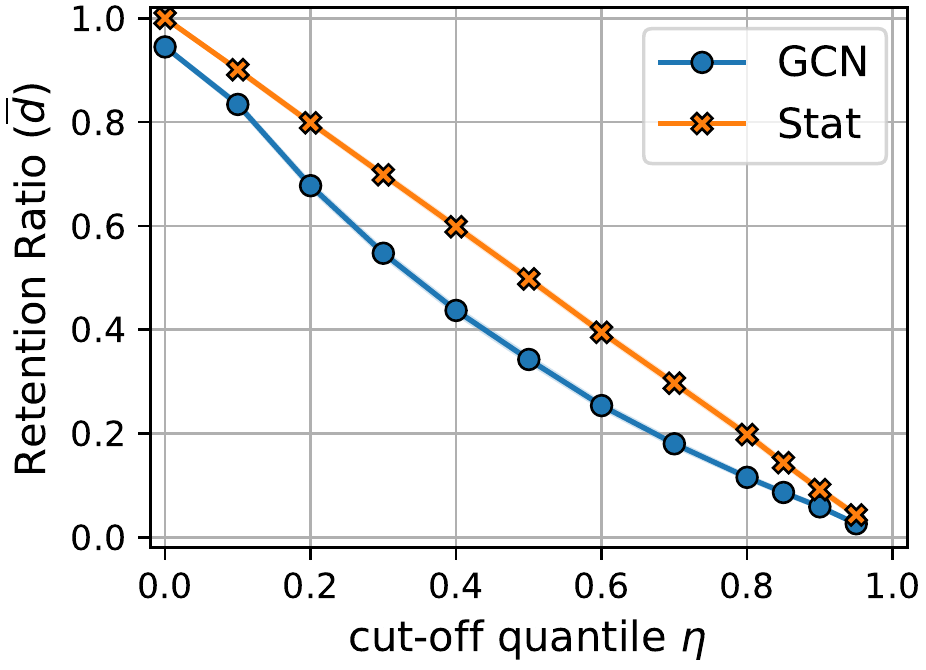}\vspace{-0.1in}
    \caption{}
    \label{fig:results:deg}
\end{subfigure}
\begin{subfigure}[b]{0.48\linewidth}
    \includegraphics[width=\linewidth]{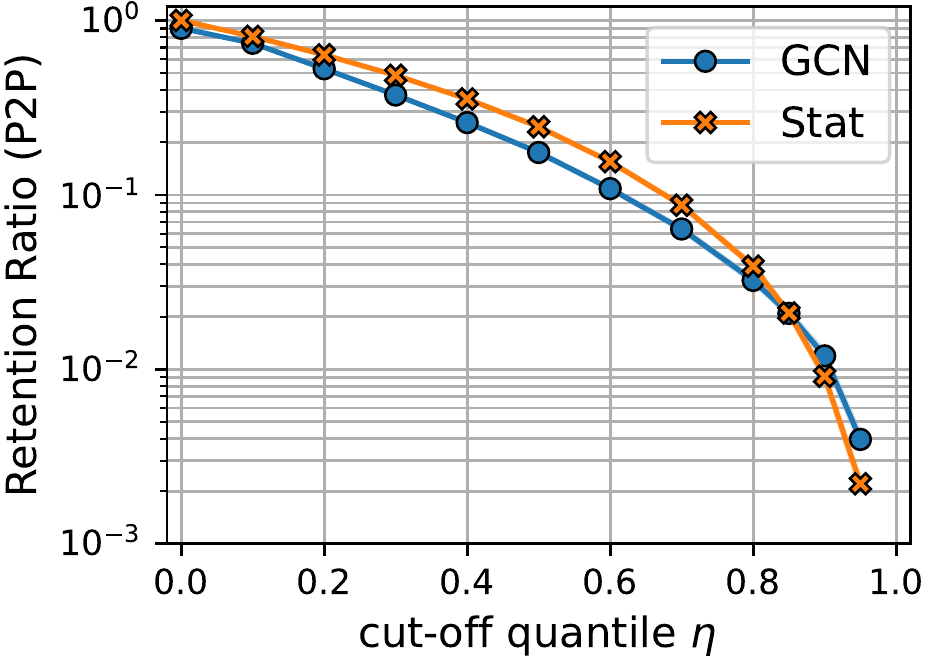}\vspace{-0.1in}
    \caption{}
    \label{fig:results:p2p}
\end{subfigure}
    \vspace{-0.12in}
    \caption{The performance by cut-off quantile $\eta$, for (a) AR for total utility, (b) RR for number of vertices in $\ccalG^s$, (c) RR for average degree of $\ccalG^s$, and (d) RR for P2P message complexity of scheduling contention, of distributed sparse schedulers w.r.t. vanilla LGS \cite{joo2012local} under identical input of $(\ccalG,\bbu)$. 
    For (c) and (d), smaller RR is better.
    }   
 \label{fig:results}    
 \vspace{-0.2in}
\end{figure}

The GCN-based link sparsification is evaluated as a component of a sparse scheduler in synthetic random conflict graphs.
The comparative baseline is the statistical link sparsification, $h_v(u(v); [\boldsymbol{1}, \boldsymbol{1}])$, described in the first paragraph of Section~\ref{sec:solution}. 
The scheduling contention function $c(\cdot)$ is selected as the local greedy solver (LGS) in \cite{joo2012local} for time-slotted networks, while the results can be easily extended to random access networks scheduled by weighted CSMA.
The performance is presented as approximation ratio (AR) or retention ratio (RR), of which the nominator is a metric of the tested sparse scheduler and the denominator is that of the vanilla LGS applied to the original graph. 
The synthetic graphs for training and testing are generated from the Erdős–Rényi (ER)~\cite{erdds1959random} model,
{which seeks to represent wireless networks with uniformly distributed users of identical omnidirectional transmit power (i.e., unit-disk interference model).}

A single layer GCN ($L=1$) is evaluated.
The training settings include a batch size of 200 for experience replay, 25 epochs, and periodic gradient reset.\footnote{Training takes 2-3 hours on a workstation with a specification of 16GB memory, 8 cores, and Geforce GTX 1070 GPU. The source code is published at \url{https://github.com/zhongyuanzhao/gcn-sparsify}}
The training set comprises 5900 random graphs drawn from the ER model, including 5000 graphs of size $V=|\ccalV|\in \left\{100, 150, 200, 250, 300\right\}$ and expected average degree $\bar{d}=Vp\in \left\{2, 5, 7.5, 10, 12.5\right\}$ ($200$ graphs per $(V, \bar{d})$), and 900 graphs of size $V\in\left\{30,100\right\}$ and probability of edge-appearance $p\in\left\{0.1,0.2,\dots,0.9\right\}$ ($50$ graphs per $(V, p)$).
The utility values are drawn from an empirical distribution (Fig.~\ref{fig:ecdf}), collected from simulation of scheduling on the synthetic conflict graph based on vanilla LGS and a utility function of $u(v)=q(v)r(v)$, where $q(v)$ and $r(v)$ are the queue length and link rate (defined as the number of packets that can be transmitted in a time slot) on link $v$. 
A 1-hop flow is generated for each node in the conflict graph. 
The exogenous packets at each source user follow a Poisson arrival with a prescribed arrival rate $\lambda$.
Our training and testing traffic load, defined as $\mu = \lambda/\mathbb{E}(\bbr)$, is set to be unsaturated as $\mu\in\left[0.03,0.05\right]$.
The link rate $r(v)$ is drawn from a normal distribution $\mathbb{N}(50, 25)$ independently across time slots and links, and clipped to $\left[0,100\right]$, 
to capture a wireless link based on constant transmit power and fading channel with lognormal shadowing \cite{Mousavi17lte}.
Our test set consists of $500$ graphs drawn from the ER model with parameters of $V\in\{100,150,\dots,300\}$, $\bar{d}\in\{2, 5, 10, 15, 20\}$, and $20$ instances per $(V, \bar{d})$.

\begin{figure}[t]
\centering
\vspace{-0.1in}
\begin{subfigure}[b]{0.47\linewidth}
    \includegraphics[width=\linewidth]{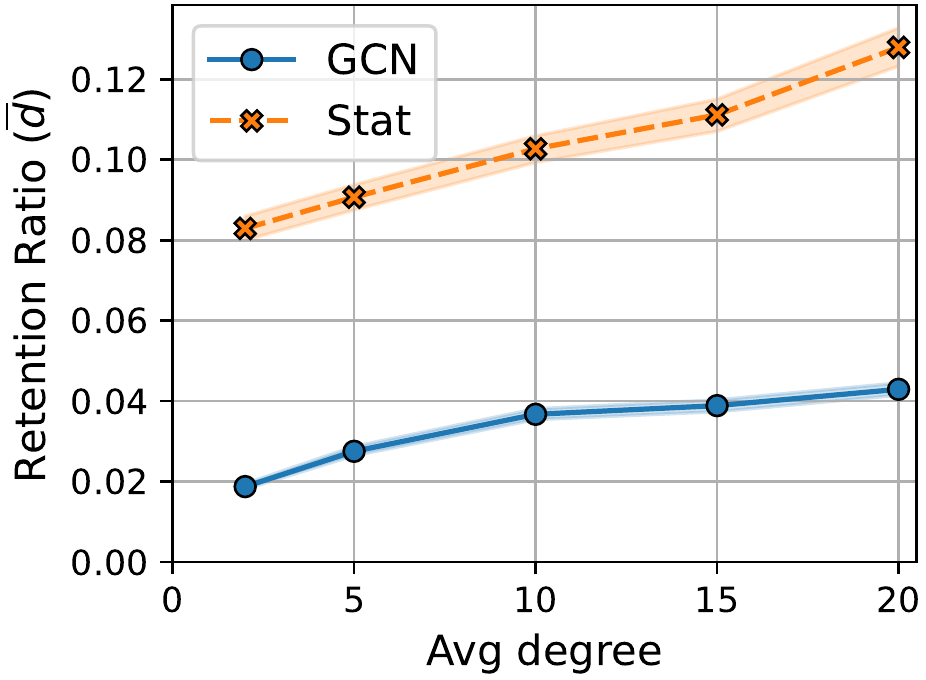}\vspace{-0.1in}
    \caption{}
    \label{fig:comm:deg}
\end{subfigure}
\begin{subfigure}[b]{0.47\linewidth}
    \includegraphics[width=\linewidth]{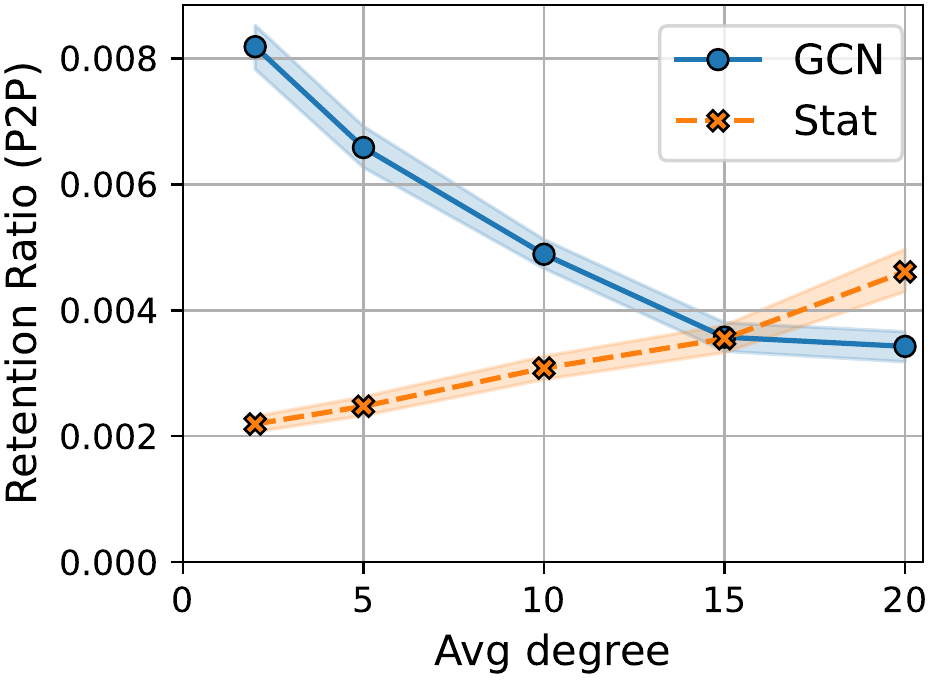}\vspace{-0.1in}
    \caption{}
    \label{fig:comm:p2p}
\end{subfigure}
    \vspace{-0.12in}
    \caption{The RRs by the average degree of original graph $\ccalG$, for (a) average degree of $\ccalG^s$, and (b) P2P message complexity of scheduling contention, of distributed sparse schedulers w.r.t. the vanilla LGS \cite{joo2012local} in 500 scheduling instances with ER graphs, traffic load $\mu\in\left[0.03,0.05\right]$, $\eta=0.95$, and $300$ time slots. Smaller RR is better.}   
 \label{fig:comm}    
 \vspace{-0.2in}
\end{figure}

Our first experiment compares the GCN-based link sparsification with the baseline under identical inputs of conflict graph and realization of random utilities.
Each graph $\ccalG(i)$ is tested with 12 cut-off quantiles $ \eta\in\{0, 0.1,\ldots, 0.8, 0.85, 0.9, 0.95\} $, 
where a realization of $\bbu$ is drawn from the collected empirical distribution for each $(\ccalG(i),\eta)$.
The performances of the GCN and the baseline are compared in Fig.~\ref{fig:results}, as the ARs of achieved total utility (Fig.~\ref{fig:results:util}), number of vertices in the sparse conflict graph $\ccalG^s$ (Fig.~\ref{fig:results:size}), average degree of $\ccalG^s$ (Fig.~\ref{fig:results:deg}), and point-to-point (P2P) message complexity of scheduling contention (Fig.~\ref{fig:results:p2p}).
Both the baseline (Stat) and GCN can significantly reduce the network-wide message complexity to less than $0.5\%$ while keep $59\sim 69\%$ of total capacity.
Compared to the baseline, GCN-based sparsification can achieve higher total utility with fewer message exchanges or collision rate (average degree of $\ccalG^s$) by leveraging the topological information. 

Next, we evaluate the GCN and the baseline in a scheduling experiment with $\eta=0.95$ and 500 scheduling instances.
Each instance contains a conflict graph from the ER test set, and realizations of random arrivals and link rates for $T=300$ time slots.
Even for the same scheduling instance, the tested schedulers generally have different input $\bbu(t)$ due to the dependency between network states and scheduling decisions.
With unsaturated traffic of $\mu\in\left[0.03,0.05\right]$, all three schedulers (vanilla LGS, GCN-based and statistical sparse schedulers) achieve the same long-term throughput.
The ARs for the average degree and P2P message complexity of scheduling contention in the sparse  graph $\ccalG^s$ are presented in Figs.~\ref{fig:comm:deg} and~\ref{fig:comm:p2p}, respectively.
The average degree of $\ccalG^s$ can be reduced to $8.2\sim12.6\%$ of that of $\ccalG$ by the baseline, and further down to $1.9 \sim 4.2 \%$ by the GCN. 
Although the baseline has lower average P2P message complexity than the GCN, as shown in Fig.~\ref{fig:results:p2p} ($\eta=0.95$) and Fig.~\ref{fig:comm:p2p}, 
their trends w.r.t. the graph density are opposite, and the GCN works better on denser graphs ($\bar{d}=20$).
These results show that GCN is superior to the baseline in reducing the scheduling overhead.

\section{Conclusions}
\label{sec:conclusions}

We presented a GCN-based distributed link sparsification scheme to reduce the scheduling overhead in wireless networks with orthogonal access. 
The GCN can encode the network topology into a parameterized local decision process to keep a link from contending if its chance of winning is low. 
The proposed scheme can significantly reduce the scheduling overhead for time-slotted and random access networks while retaining most of the network capacity. 
Thus, our approach can improve   access, energy efficiency, and radio footprint of wireless multi-hop networks.


\vfill\pagebreak



\bibliographystyle{ieeetr}
\bibliography{strings,refs}

\end{document}